\documentstyle[12pt]{article}

\begin{document}


\begin{center}
\noindent
{\Large \bf The Evolution Equation of Cosmological Density Perturbation in Einstein-Cartan Gravity}\\[3mm]
by \\[0.3cm]
{\sl L.C. Garcia de Andrade\footnote{Electronic mail: garcia@symbcomp.uerj.br}}
\\[0.3cm]
Departamento de F\'{\i}sica Te\'orica\\[-3mm]
Universidade do Estado do Rio de Janeiro -- UERJ\\[-3mm]
Cep 20550, Rio de Janeiro, RJ, Brasil   
\vspace{2cm}

{\bf Abstract}
\end{center}
\paragraph*{}

The evolution equation of linear cosmic density perturbations in the realm of Einstein-Cartan theory of gravity is obtained.The de Sitter metric fluctuation is computed in terms of the spin-torsion background density.

\newpage

Investigation of linear density fluctuations \cite{1,2} in the context of general relativistic cosmology have proved to be useful in the study of structure formation like galaxy formation for example.More recently D.Palle \cite{3} have investigated the Primordial density fluctuations in Einstein-Cartan cosmology and made comparison with the COBE data.Later on we developed \cite{4} some of his ideas to solutions of the Einstein-Cartan cosmology with inflatons and dilatons.In this letter we show that
the evolution of density fluctuations can be derived from the Einstein-Cartan field equations and the associated conservation equation for the matter and spin-torsion density in analogous way that it is done in General Relativity as long as some simple assumptions are made on the galatic spinning fluid.A stationary metric here is not needed since we are not considering that the spin of the intrinsic particles affect appreciably the rotation of galaxies.Therefore the usual Friedmann metric to investigate linear perturbations in Einstein-Cartan gravity.Let us begin by considering the Friedmann metric
\begin{equation}
ds^{2}=dt^{2}-a^{2}(dx^{2}+dy^{2}+dz^{2})
\label{1}
\end{equation}   
where $a(t)$ is the cosmic scale.The Einstein-Cartan equations are given by 
\begin{equation}
H^{2}=\frac{8{\pi}G}{3}({\rho}_{eff}-2{\pi}G{\sigma}^{2})
\label{2}
\end{equation}
and 
\begin{equation}
H^{2}+\dot{H}=-\frac{4{\pi}G}{3}({\rho}_{eff}-8{\pi}G{\sigma}^{2})
\label{3}
\end{equation}
where $H=\frac{\dot{a}}{a}$ is the Hubble parameter in terms of time.Where in short we use the following notations
\begin{equation}
{\rho}_{eff}={\dot{\phi}}^{2}+V
\label{4}
\end{equation}
and
\begin{equation}
p_{eff}={\dot{\phi}}^{2}-V
\label{5}
\end{equation}
where ${\phi}$ is the inflaton field ,$V$ is the inflaton potential and ${\sigma}^{2}=<S_{ij}S^{ij}>$ is the averaged squared of the spin density tensor $S_{ij}$.  
By making use of the definition ${\rho}={\rho}_{eff}-2{\pi}G{\sigma}^{2}$ and $p=p_{eff}-2{\pi}G{\sigma}^{2}$ allow us to write the conservation equation in Einstein-Cartan Gravity \cite{4} as
\begin{equation}
\frac{d}{dR}({{\rho}R^{3}})=-3pR^{2}
\label{6}
\end{equation}
These substitutions allow us to reduce our problem of computing density perturbations formally similar to the general relativistic one.The necessary assumption to make our task easier is to consider the pressure $p$ vanishes 
which yields
\begin{equation}
p_{eff}=2{\pi}G{\sigma}^{2}
\label{7}
\end{equation}
With these assumptions equation \ref{6} reduces to
\begin{equation}
\frac{d}{dt}({{\rho}R^{3}})=0
\label{8}
\end{equation}
where now we are ready to apply the traditional perturbation method on the
cosmological density perturbation by making use of the following definitions
\begin{equation}
\delta=\frac{{\rho}-{\rho}_{b}}{{\rho}_{b}}
\label{9}
\end{equation}
and
\begin{equation}
a=R(t)+{\delta}R(t)
\label{10}
\end{equation}
and
\begin{equation}
{\sigma}^{2}={\sigma}^{2}_{b}(1+\frac{{\delta}{\sigma}^{2}}{{\sigma}^{2}})
\label{11}
\end{equation}
Here the index b denotes background quantities.From the equation (\ref{8}) one obtains
\begin{equation}
{\delta}=-3\frac{{\delta}R}{R}
\label{12}
\end{equation}
following the lines and procedures of General Relativity we find the following evolution equation
yields
\begin{equation}
{\ddot{\delta}}+2{H_{0}}^{2}{\dot{\delta}}+10{\pi}^{2}G^{2}{{\sigma}_{b}}^{2}{\delta}=0
\label{13}
\end{equation}
To solve this differential equation we need to compute the background spin-torsion density.This can be easily accomplished if one equates equations (\ref{2}) and (ref{3}) and substitute (\ref{7}) to obtain 
\begin{equation}
{\rho}_{eff}=\frac{9}{2}{\pi}G{{\sigma}_{b}}^{2}
\label{14}
\end{equation}
Substitution of this last expression into the conservation equation for the effective density
yields
\begin{equation}
{\dot{{{\sigma}_{b}}^{2}}}+3H_{0}{{\sigma}_{b}}^{2}=0
\label{15}
\end{equation}
which solution is
\begin{equation}
{{\sigma}_{b}}^{2}=e^{-3H_{0}t}
\label{16}
\end{equation}
to simplify matters without loosing too much physical insight one could assume that the spin-torsion background can be expanded in the form
\begin{equation}
{{\sigma}_{b}}^{2}={(1+3H_{0}t)}^{-1}={(3H_{0}t)}^{-1}
\label{17}
\end{equation}
where we have made the hypothesis $H_{0}t>>\frac{1}{3}$.Substitution of this result into expression (\ref{13}) yields the final form of the evolution equation ready now to be solved 
\begin{equation}
{\ddot{\delta}}+2{\alpha}{\dot{\delta}}+{\gamma}\frac{{\delta}}{t}=0
\label{18}
\end{equation}
where ${\alpha}=2{H_{0}}^{2}$ and ${\gamma}=\frac{-10{\pi}G}{3H_{0}}$.To deduce the evolution equation (\ref{18}) we assumed that the perturbed metric is the de Sitter metric where 
$H_{0}$ is a constant.To solve this evolution equation let us assume as usual a solution of the type ${\delta}=t^{m}$ where $m$ is a real constant to be determined from an algebraic equation.Substitution of this hint into the differential  equation (\ref{18}) yields the following algebraic equation
\begin{equation}
m^{2}+(1+{\alpha})m+{\gamma}=0
\label{19}
\end{equation}
the solution of this last equation left us with two solutions of the type ${\delta}_{+}=t^{m_{+}}$ and $ {\delta}_{-}=t^{m_{-}}$ where $m_{+}$ and $m_{-}$ are the respective elementary solutions of the algebraic equation (ref{19}). 
Recently Maroto and Shapiro \cite{5} have discussed the stability of de Sitter metric on a higher order derivative gravity with torsion.They basically showed that the stability of de Sitter metric depends on the dimension of the space with torsion we consider therefore it is interesting to see what happens here in $4D$ spacetime.Here we show that is possible to extract important physical information on the stability of de Sitter metric by using the perturbation method discussed in the last section.To this aim we consider the following perturbation on the de Sitter metric
\begin{equation}
H=H_{0}+{\delta}{H}
\label{20}
\end{equation}
Substitution of this equation into the Friedmann equation after some algebra yields to the first order
\begin{equation}
{\delta}H=\frac{{\rho}_{b}-2{\pi}G{{\sigma}_{b}}}{H_{0}t^{m}}=\frac{5{\pi}G}{2{H_{0}}^{2}t^{m+1}}
\label{21}
\end{equation}
This expression can be recast in the form
\begin{equation} {\delta}H=\frac{\frac{5}{2}{\pi}G{{\sigma}_{b}}}{H_{0}t^{m}}=\frac{5{\pi}G}{2{H_{0}}^{2}t^{m+1}}
\label{22}
\end{equation}
which shows that the perturbation of the de Sitter metric will depend explicitly on the background spin-torsion density.
\paragraph*{} 
\vspace{1cm}
\noindent
{\large\bf \underline{Acknowledgements}}: I am very much indebt to
Professor Ilya Shapiro for his constant advice.Financial support from CNPq (Brazilian Government Agency) and UERJ are gratefully acknowledged.

\end{document}